\documentclass[11pt]{article}
\usepackage{rotating}
\usepackage{amsmath,amssymb}
\usepackage{graphicx}
\usepackage[utf8]{inputenc}
\DeclareUnicodeCharacter{2212}{-}
\usepackage{authblk} 
\usepackage{xcolor}
\usepackage{graphicx}
\graphicspath{{graphs/}}
\usepackage{amsmath,amssymb}
\usepackage[colorlinks=true, linkcolor=black, citecolor=blue, urlcolor=blue]{hyperref}

\usepackage{subcaption}
\usepackage[utf8]{inputenc}   
\usepackage[T1]{fontenc}      
\usepackage{textcomp}      
\usepackage{natbib}
\usepackage{listings}
\usepackage{subcaption}
\usepackage{tocloft}
\usepackage{multirow}
\usepackage{float}

\UseRawInputEncoding
\begin{document}
\title{Revisiting 2D and 3D Dainotti Correlations for GRBs Using Bayesian Neural Networks}
\author[1]{Nilanjana Bagchi Aurpa \thanks{Email : {\url{nbagchi01@gmail.com}}}}
\author[2]{Abha Dev Habib \thanks{Email : {\url{abhadev.habib@mirandahouse.ac.in}}}}
\author[2]{Nisha Rani \thanks{Email : {\url{nisha.physics@mirandahouse.ac.in}}}}
\affil[1]{Universit\'e Paris Cit\'e, F-75013 Paris, France}
\affil[2]{Miranda House, University of Delhi, Delhi 110007, India}

\maketitle
\begin{abstract}
\noindent Gamma-ray bursts (GRBs) are promising cosmological probes, but their use as standard candles is limited by the circularity problem, necessitating model-independent calibration of GRB luminosity correlations. We revisit the two-dimensional (2D) and three-dimensional (3D) Dainotti correlations using Bayesian Neural Networks (BNNs) trained on the updated Observational Hubble Data (OHD) and Pantheon+ Type Ia Supernova sample. The reconstructed luminosity distances are used to calibrate the Platinum and Narendra \textit{et al.} GRB samples. We constrain the parameters of the 2D Dainotti relation and the 3D fundamental plane, and examine the impact of calibration datasets and GRB sample selection. Calibration achieved using Pantheon+ yields tighter constraints than OHD, while the 3D correlation exhibits lower intrinsic scatter than the 2D relation. Our results demonstrate that BNNs provide a robust framework for model-independent calibration of GRB luminosity correlations with reliable uncertainty propagation. Further, the underlying distance probe is a key factor in model-independent calibration, determining both the size of the GRB samples and the precision of the resulting constraints.
\end{abstract}
\section{Introduction}
Gamma-Ray Bursts are one of the most luminous transient events in the universe. Due to their extraordinary luminosities, GRBs have been observed up to $z \sim 9.4$ \cite{Salvaterra2009,Cucchiara2011}. Their ability to probe the high-redshift universe far beyond the reach of Supernovae Ia makes them valuable cosmological distance indicators.  The theoretical predictions suggest that they could be detectable out to $z \sim 20$ \cite{Lamb2002}. Consequently, improving the calibration of GRB luminosity correlations is crucial for establishing GRBs as reliable standardizable candles and for extending cosmology to the earliest epochs of the universe.\\

\noindent Several empirical correlations have been proposed in the attempt to establish GRBs as standardizable candles \cite{Amati2002,Ghirlanda:2004fs,Yonetoku2003,Izzo:2015vya,Liang2005}. Some notable examples include the Amati relation, Yonetoku relation, Dainotti relation and Ghirlanda relation. Amati relation relates the rest-frame spectral peak energy, $E_{\mathrm{p}}$, with the isotropic-equivalent radiated energy, $E_{\mathrm{iso}}$ \cite{Amati2002, Amati:2008hq}. Yonetoku relation links $E_{\mathrm{p}}$ to the peak luminosity, $L_{\mathrm{p}}$ \cite{Yonetoku2003}. The Ghirlanda correlation is a remarkably tight empirical relation that connects $E_{\mathrm{p}}$ of the prompt emission with the collimation-corrected jet energy, $E_{\gamma }$ \cite{Ghirlanda:2004me}. The luminosity correlation of the plateau, commonly known as the Dainotti relation, connects the luminosity at the end of the X-ray plateau phase and its rest-frame duration \citep{Dainotti2008,Dainotti2011,
Dainotti2013,Dainotti2016,Dainotti2018,
Dainotti2021}. \\

\noindent Here, we restrict our study to the Dainotti relation because this relation possesses several advantages. For example, it is supported by a plausible physical interpretation based on magnetar emission \cite{Rea2015,Rowlinson2014,Bernardini2015}. In addition, its calibration has been shown to remain robust after accounting for selection effects \cite{Dainotti2016Sne,Dainotti2013}. Furthermore, it is believed that the prompt-afterglow correlations are particularly promising for standardizing GRBs as they generally exhibit lower intrinsic scatter than those which are solely based on prompt emission characteristics \cite{Cardone2010}. Intrinsic scatter represents variance not explained by observational uncertainties. The two-dimensional (2D) Dainotti correlation has been extensively employed as a cosmological distance indicator \cite{Cardone2010,Dainotti2013}. The three-dimensional (3D) Dainotti relation, which relates the prompt peak luminosity, time of the rest-frame plateau, and the corresponding plateau luminosity \cite{Dainotti2016,Dainotti2017fp} further reduces intrinsic scatter. \\

\noindent As GRB luminosity distance, $d_L$ is not an observable, one frequently used approach is to use fiducial cosmological model to  first calibrate GRB luminosity distances, and then utilize them to constrain alternate models. Doing so gives rise to a major issue, termed as the circularity problem \cite{Wang:2015ira,Ghisellini2005,Ghirlanda:2006ax}. As mentioned, such an approach introduces a model-dependent bias limiting the reliability of the inferred cosmological parameters. \\

\noindent For breaking the circularity problem, a wide range of methods have been developed for the calibration of the GRBs using observational datasets, including interpolation-based techniques \cite{Liang2008,Liu:2022inf}, local regression schemes \citep{Cardone:2009mr,Demianski:2016zxi,Demianski:2019vzl}, Bezier parametric reconstructions \citep{Amati2019}, iterative calibration procedures \citep{Liang2008} and approximations \cite{Liu:2014vda}. Gaussian Process regression has also emerged as a widely used tool for model-independent cosmological reconstruction and represents one of the earliest applications of machine learning techniques in this field \cite{Seikel2012a,Seikel:2012cs,Li:2017zrx,Han2024,Pan:2020zbl,Mu:2023bsf,Mu:2023zct,Sun:2021pbu,Zhang:2024ndc,Kumar:2022ypo,Favale:2024lgp}. However, due to the sensitivity of Gaussian Processes to the choice of kernel, the reliability of the reconstructed functions may be affected \cite{Zhang:2023xgr,Wei:2016xti, Johnson2025blf}. Artificial Neural Networks (ANN) are inherently more data driven and impose significantly fewer assumptions on the properties of the data. There have been previous studies exploring their utility for cosmological applications \cite{Zhang:2024ndc,Dialektopoulos:2021wde,Wang2019, Aurpa2026}.   However, ANNs do not inherently provide uncertainty estimates for their predictions, which raises concerns about the reliability of predictions for observational datasets with large uncertainties, such as the Observational Hubble Data (OHD) \cite{Mahida:2025teg}. \\

\noindent For model-independent constraints on Dainotti Relations different methods have been explored. Cao et al. 2022 \cite{Cao2022s} simultaneously fitted cosmological constraints with Dainotti Parameters. Favale et al.(2024) \cite{Favale:2024lgp} studied the model independent correlations using Gaussian Process, followed by Mukherjee et al. 2025 \cite{Mukherjee:2024akt} using ANN with Pantheon+ dataset using full covariance matrix. \\

\noindent In our study, we employ Bayesian Neural Network (BNN) to obtain the luminosity distance from observational datasets. In order to reduce model bias in the predicted observable, BNN utilises  Bayesian Inference to provide uncertainty in model parameters \citep{bishop2006prml,bishop2013brml,Gelman2013,ghahramani2015probabilistic,Neal1996}.  
Here, we study the effects of different calibration  datasets on the Dainotti correlation parameters. Following the reconstruction, we calibrate two different GRB samples the Platinum \cite{Cao2022s} and the latest collection by Narendra et al. \cite{Narendra2024}. Finally, we constrain the parameters of 2D and 3D Dainotti relations.\\

\noindent The paper is structured as follows - in Section \ref{bnn}, we discuss the formulation of BNN and architectural choices. We reconstruct OHD and Pantheon+ data using BNN in Section \ref{ohdbnn} and \ref{Snebnn} respectively. In Section \ref{2ddai} and \ref{3ddai}, we constrain 2D and 3D Dainotti correlations. We compare our findings with previous studies in Section \ref{result}. We conclude our paper with a discussion in Section \ref{conclu}. 
\section{Data and Methodology}

\subsection{Training Datasets}

Like all neural networks, BNN is a data driven framework. As different observational datasets are populated differently across redshifts, we reconstruct the luminosity distance using two independent training datasets to investigate the Dainotti relations. We first construct a BNN model for the updated Observational Hubble Data (OHD) from Table~1 of Ratra et al.~\cite{Ratra2023}, hereafter referred to as \textbf{BNN-OHD}. The dataset comprises 32 data points spanning the redshift range $0.07 < z < 1.965$.\\

\noindent We construct another BNN model trained over the Pantheon+ Type Ia Supernovae (SNe Ia) sample, hereafter referred to as \textbf{BNN-Pantheon}. The Pantheon+ dataset extends to a redshift of $z\sim2.26$\footnote{\url{https://github.com/PantheonPlusSH0ES/DataRelease}} and has been widely used as a standard candle for model-independent calibration in previous studies \cite{Han2024, Huang2025p,Shah2024}. It consists 1701 SNe Ia light curves.
\subsection{GRB Samples}
The calibration of the Dainotti relation is known to depend on the selection criteria and the characteristics of the underlying GRB sample. To investigate the impact of sample selection and dataset size on the inferred correlation, we employ two independent GRB datasets with different levels of homogeneity and completeness    \cite{Dainotti2021juc}.\\

\noindent Further, the redshift range of the training dataset decides the subset of GRB sample that can be used to constrain the parameters of GRB correlations. In order to understand the effect of selection criteria and strength of GRB dataset on the constraints on the Dainotti relation, we proceed with two different GRB
datasets.\\

\noindent We first consider the Platinum Dainotti sample compiled by Cao et al.~\cite{Cao2022s}. This sample consists of 50 long GRBs ($T_{90}>2\,\mathrm{s}$) in the redshift range $0.553 \leq z \leq 5.0$, selected for their well-defined X-ray plateau emission. Specifically, the sample includes GRBs with plateau inclinations smaller than $41^\circ$, plateau durations longer than $500\,\mathrm{s}$, and no X-ray flares, resulting in the lowest intrinsic scatter in the Dainotti relation and making them suitable for cosmological standardization. Dictated by the redshift overlap, a sub-sample of 19 GRBs is used for constraining Dainotti relations by using calibration obtained from the BNN-OHD while the sample size increases to 26 GRBs when we use calibration from the BNN-Platinum.\\

\begin{figure}[h!]
    \centering
    \includegraphics[width=0.8\textwidth]{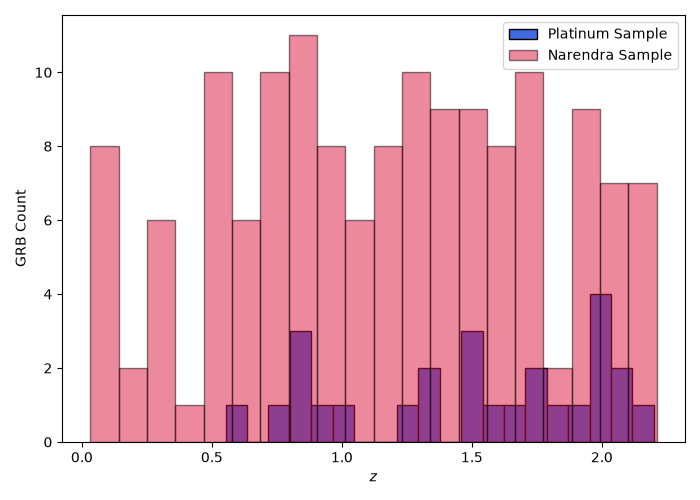}
    \caption{Redshift distribution of the sub-samples of the Platinum and Narendra Sample, which are used to constrain Dainotti Relations using calibration from the BNN-Pantheon till z$\sim$2.26. }
    \label{fig:hist}
\end{figure}
  
\noindent Selection effects are known to influence GRB luminosity correlations, since observational samples may not fully represent the underlying GRB population. The Platinum sample was designed to minimize such biases through stringent selection criteria, resulting in a low-scatter sample suitable for cosmological standardization.\cite{Dainotti:2017fhk}\\

\noindent To examine the robustness of the calibrated relations beyond a highly selected sample, we further employ the catalogue compiled by Narendra et al.~\cite{Narendra2024}. The catalogue\footnote{\url{https://github.com/gammarayapp/GRB-Web-App}} provides a comprehensive and homogeneous compilation of 246 long GRBs with uniformly collected prompt-emission and X-ray afterglow observables derived primarily from the Swift observations, covering a redshift range of $0.0331 \leq z \leq 8.26$. The larger and more representative sample enables us to investigate the stability of the 2D and 3D Dainotti relations beyond the low-scatter Platinum sample. From the dataset provided by Narendra et al.~\cite{Narendra2024}, a subset of 131 GRBs are used for the OHD calibration, while a subset of 147 GRBs are used to constrain the 2D and 3D Dainotti relations using calibrations obtained from the BNN-Pantheon.
\begin{table}[ht]
\centering
\label{tab:grb_samples}
\begin{tabular}{l|c|c|c}
\hline
Dataset & Total GRBs & OHD Calibration & Pantheon+ Calibration \\
\hline
Platinum Sample \citep{Cao2022s} & 50 & 19 & 26 \\
Narendra et al. \citep{Narendra2024} & 246 & 131 & 147 \\
\hline
\end{tabular}
\caption{Summary of the GRB datasets used in this work and the number of GRBs employed for calibration using the BNN-OHD and BNN-Pantheon.}
\end{table}

\subsection{Reconstruction using BNN}
\label{bnn}
A Bayesian Neural Network utilises Bayesian 
Inference to optimise the model parameters providing aleatoric and epistemic uncertainty \citep{Neal1996,MacKay1995,Gelman2013,
ghahramani2015probabilistic}. Given the observed data $\mathcal{D}$, the posterior distribution of the parameters is obtained via Bayes' theorem,
\begin{equation}
p(\mathbf{w} \mid \mathcal{D}) = \frac{p(\mathcal{D} \mid \mathbf{w})\, p(\mathbf{w})}{p(\mathcal{D})},
\end{equation}
where $p(\mathcal{D} \mid \mathbf{w})$ denotes the likelihood and $p(\mathcal{D})$ is the probability of observing the data. $p(\mathrm{w})$ denotes the prior of the weights. \\

\noindent The assumption of Gaussian uncertainties may not be valid for heterogeneous or high-redshift datasets \citep{dainotti2024,Favale:2024lgp}. By marginalizing over the posterior distribution of network parameters, the BNN framework provides a more robust treatment of uncertainty.
Following the assumption, the likelihood function is written as
\begin{equation}
p(\mathbf{y} \mid \mathbf{x}, \mathbf{w}) \propto 
\exp\left[
-\frac{1}{2}\sum_i
\frac{\left(y_i - f(\mathbf{x}_i;\mathbf{w})\right)^2}{\sigma_i^2}
\right],
\end{equation}
where $f(\mathbf{x};\mathbf{w})$ denotes the network output and ($x_i$, $y_i$) are elements of the dataset D. \\

\noindent The No-U-Turn Sampler (NUTS) \cite{hoffman2011}, an adaptive variant of Hamiltonian Monte Carlo (HMC) is used as supplement to MCMC to infer posterior distribution of network parameters. The NUTS algorithm  adaptively selects parameters during the warm-up phase which automates the sampling procedure and improves sampling efficiency. The final prediction is obtained by marginalizing the likelihood.\\

\noindent The choice of BNN model requires optimisation of different hyperparameters - hidden layers, number of neurons, choice of prior. Following the approach by Aurpa et al. \cite{Aurpa2026} the comparison across the hyperparameter configurations is performed using the Widely Applicable Information Criterion (WAIC)  \cite{watanabe2010asymptotic,vehtari2017practical}.  WAIC is computed for each combination of network width and prior variance to evaluate the balance between predictive accuracy and model flexibility. \\

\subsubsection{Observational Hubble Data}
\label{ohdbnn}
\begin{figure}[h!]
    \centering
    \includegraphics[width=0.8\textwidth]{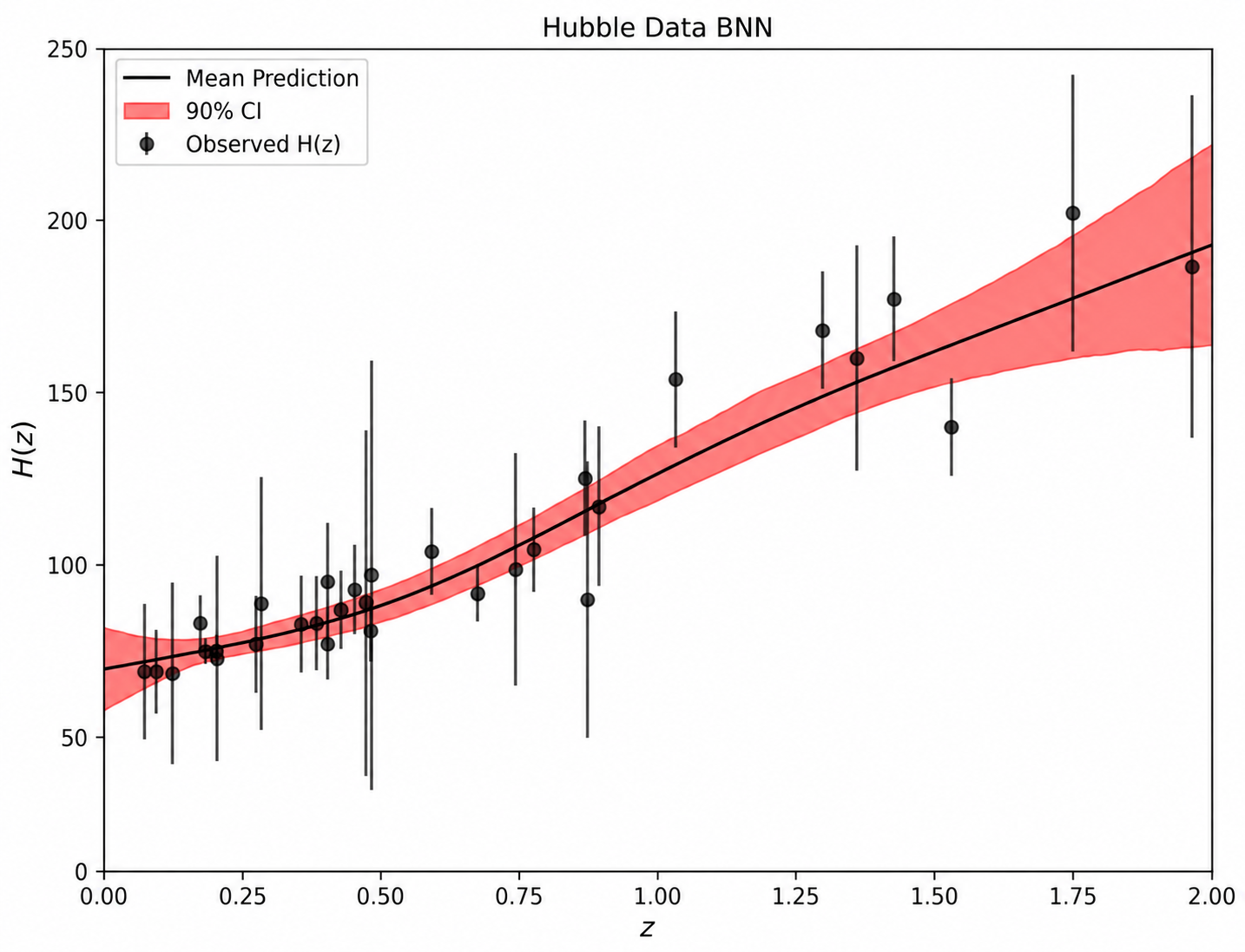}
    \caption{Bayesian Neural Network reconstruction of the OHD Hubble Parameter $H(z)$, as a function of redshift. The black line illustrates predictions obtained from the BNN-OHD with the associated 1 $\sigma$ uncertainty in red. }
    \label{fig:hdbnn}
\end{figure}  
\noindent Following the stated approach, we perform grid search and compare WAIC criterion for different models to obtain the optimum BNN model for OHD reconstruction. The bfit results are summarized in Table \ref{tab:bnnmodelOHD}. \\

\begin{table}[h!]
\centering
\begin{tabular}{lc}
\hline
Hyperparameter & Value \\
\hline
Hidden layers & 1 \\
Neurons per layer & 64 \\
Prior &(0,$5^2$)\\
Activation function & ELU \\
\hline
\end{tabular}
\caption{Optimum Parameters for the BNN-OHD. We borrow the architecture as used in Aurpa et al. \cite{Aurpa2026}. }
\label{tab:bnnmodelOHD}
\end{table}

\noindent The reconstruction of the OHD using the BNN framework is illustrated in Figure \ref{fig:hdbnn}. FRom Figure 2, we see that the BNN successfully captures the observed evolution of the Hubble parameter over the full redshift range of the OHD sample. The mean prediction closely follows the measurements, while the shaded credible region represents the predictive uncertainty obtained by marginalizing over the posterior distribution of the network parameters. The uncertainty remains relatively small in the densely sampled redshift region and gradually increases toward higher redshifts where the observational constraints become sparse.\\

\noindent The luminosity distance, $d_L(z)$ is obtained by integrating the inverse expansion rate $H(z)$ : 
\begin{equation}
d_L(z) = (1+z)\, c \int_0^z \frac{\mathrm{d}z'}{H(z')},
\end{equation}
where $c$ is the speed of light.

\subsubsection{Pantheon+ Data}
\label{Snebnn}
\begin{figure}[h]
\centering
\includegraphics[width=0.8\textwidth]{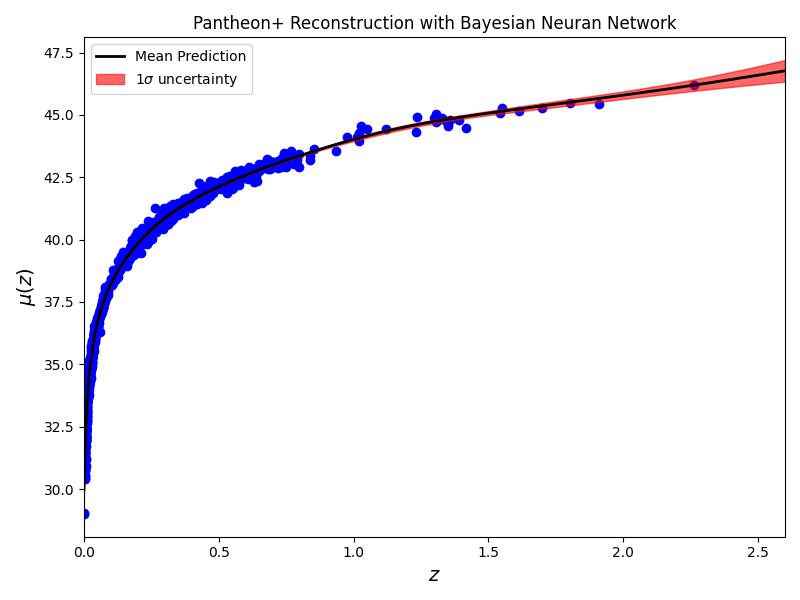}
\caption{BNN-Pantheon reconstruction of distance modulus, $\mu(z)$, as a function of redshift. The blue points represent the Pantheon+ SNe Ia observations, the solid black curve denotes the mean BNN reconstruction, and the shaded red region corresponds to the $1\sigma$ predictive uncertainty obtained from the posterior distribution of the network parameters. Although the reconstruction extends over the full Pantheon+ redshift range, only the reconstructed distance modulus up to $z=2.26$ is used for the calibration of the GRB luminosity correlations. }
        \label{fig:panbnn}
\end{figure}    
 \noindent We similarly perform a grid search to obtain the optimum parameters for the model for reconstruction of Pantheon+. The optimum parameters for the model BNN-Pantheon is shown in Table \ref{tab:panbnnmodel}. Following the reconstruction of the Pantheon+ dataset, we calibrate GRBs using the framework. \\
\begin{table}[h]
\centering
\begin{tabular}{lc}
\hline
Parameter & Value \\
\hline
Hidden layers & 1 \\
Neurons per layer & 64 \\
Prior &(0,$10$)\\
Activation function & ELU \\
\hline
\end{tabular}
\caption{Optimum Parameters for constructing the BNN-Pantheon.}
\label{tab:panbnnmodel}
\end{table}

\noindent The distance modulus can be utilised to obtain the luminosity distance and related uncertainty as:
\begin{equation}
d_L = 10^{\frac{\mu-25}{5}}\ \mathrm{Mpc}.
\end{equation}
\begin{equation}
\sigma_{d_L}
=
\frac{\ln 10}{5}\,
d_L\,\sigma_\mu.
\end{equation}
We constrain both the 2D Dainotti Relations and 3D Dainotti Relations in the subsequent sections.
\begin{figure}[h!]
    \centering
    \includegraphics[width=0.8\textwidth]{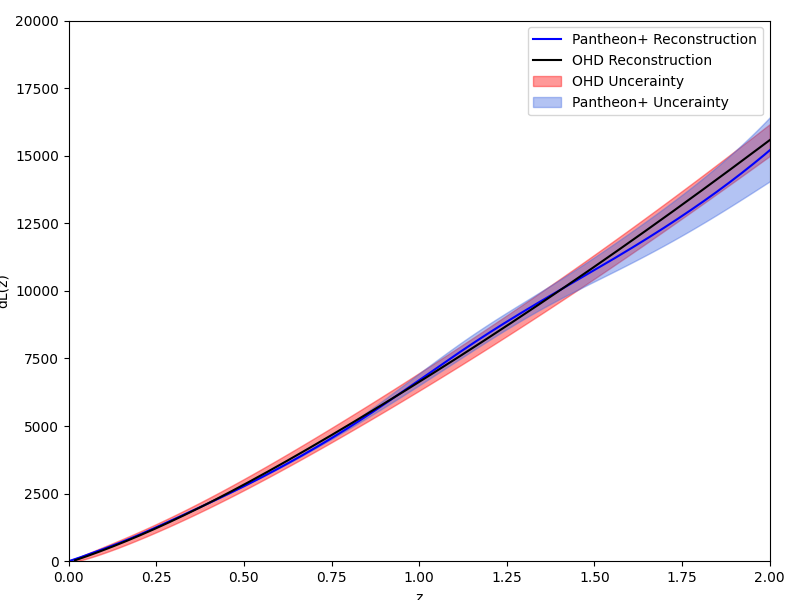}
    \caption{BNN-OHD predicts till redshift range z $\sim$ 2 while the BNN-Pantheon extends up to z$\sim$2.26.  We compare the reconstructed luminosity distance, $d_L(z)$, obtained using the BNN-OHD and BNN-Pantheon in the common redshift range. The black and blue curves are reconstructions obtained from the BNN-OHD and BNN-Pantheon respectively, while the red and blue shaded regions denote their corresponding $1\sigma$ uncertainty bands. The two reconstructions show excellent agreement over the redshift range $0 \leq z \leq 2$, with only minor differences as z approaches 2 . }
        \label{fig:combobnn}
\end{figure}    

\subsection{Constraints on the 2D Dainotti Correlation}
\label{2ddai}
Following the calculation of luminosity distance, the X-ray source rest-frame luminosity can be obtained as: 
\begin{equation}
L_{X}=4\pi d_{L}^{2}F_{X}K_{\rm plateau},
\end{equation}
The associated uncertainty can be obtained as-
\begin{equation}
\sigma_{L_X} = L_X
\sqrt{
\left(\frac{2\sigma_{d_L}}{d_L}\right)^2
+
\left(\frac{\sigma_{F_X}}{F_X}\right)^2
+
\left(\frac{\sigma_{K_{\rm plateau}}}{K_{\rm plateau}}\right)^2
}.
\end{equation}
The luminosity is expressed in units of
$\mathrm{erg\,s^{-1}}$. The $F_{X}$ is the measured X-ray energy flux at the
end of the plateau in
units of $\mathrm{erg\,cm^{-2}\,s^{-1}}$. $K_{\rm plateau}$ is the
log-power plateau $K$-correction, with $\alpha_{\rm plateau}$ as the
X-ray photon index of the plateau phase. The prompt-emission
$K$-correction and  $K_{\rm prompt}$ are obtained from the spectral parameters.\\

\noindent The 2D Dainotti correlation is given as:
\begin{equation}
\log L_X = C_0 + a\log T_a^{*}
\end{equation}

\noindent Here, $T_{a}^{*}$ refers to the characteristic rest-frame
time corresponding to the end of the plateau emission. 
The likelihood is given by:
\begin{equation}
\ln \mathcal{L}_{\rm 2D}
=
-\frac{1}{2}
\sum_{i=1}^{N}
\left[
\frac{
\left(
\log L_{X,i}
-
C_0
-
a\log T_{a,i}^{*}
\right)^2
}
{\sigma_i^2}
+
\ln(2\pi\sigma_i^2)
\right],
\end{equation}
where - 
\begin{equation}
\sigma_i^2
=
\sigma_{\log L_{X,i}}^2
+
a^2\sigma_{\log T_{a,i}^{*}}^2
+
\sigma_{\rm int}^2.
\end{equation}

\noindent We adopted weakly informative uniform priors for all free parameters, with bounds chosen to encompass the physically plausible parameter space while minimizing the influence of prior assumptions on the posterior distributions. A constraint of $a<0$ was imposed for consistency with physical interpretation while the intrinsic scatter was restricted to $\sigma_{\rm int}>0$ by definition.\\

\begin{figure}[H]
    \centering
    \includegraphics[width=0.7\textwidth]{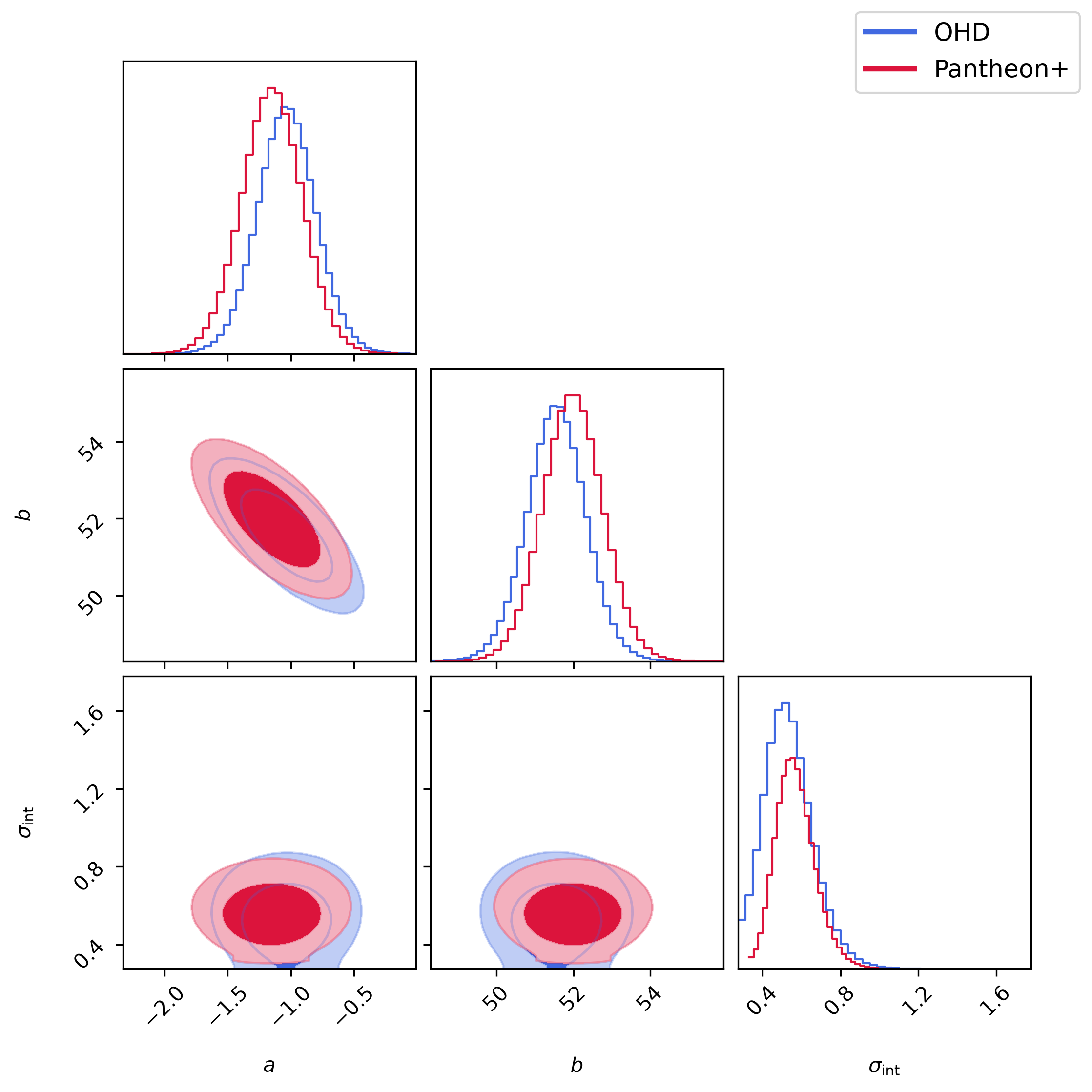}
    \caption{Corner plot showing the posterior distributions of the marginalized 1$\sigma$ and 2$\sigma$ credible intervals of the 2D Dainotti relation parameters $(a,\,C_0,\,\sigma_{\rm int})$. The red contours correspond to the GRB subset
which is calibrated
using the BNN-Pantheon
while the blue
is used for constraints
obtained from the subset
which uses calibrations
obtained from the BNN-OHD.}
    \label{fig:2dplat}
\end{figure}

\begin{table}[h]
\centering

\begin{tabular}{lcc}
\hline
Parameter & 2D Dainotti & 3D Fundamental Plane \\
\hline
$a$ & $\mathcal{U}(-5,\,0)$ & $\mathcal{U}(-5,\,5)$ \\
$b$ & --- & $\mathcal{U}(0,\,5)$ \\
$C_0$ & $\mathcal{U}(0,\,100)$ & $\mathcal{U}(0,\,100)$ \\
$\sigma_{\rm int}$ & $\mathcal{U}(0,\,5)$ & $\mathcal{U}(0,\,5)$ \\
\hline
\end{tabular}
\caption{Uniform priors adopted for the Bayesian MCMC analysis of the Dainotti 2D correlation and 3D Fundamental Plane.}
\label{tab:priors}
\end{table}
\subsection{Constraints on 3D Fundamental Plane}
\label{3ddai}

\begin{figure}[ht]
    \centering

    \begin{subfigure}{0.48\linewidth}
        \centering
        \includegraphics[width=\linewidth]{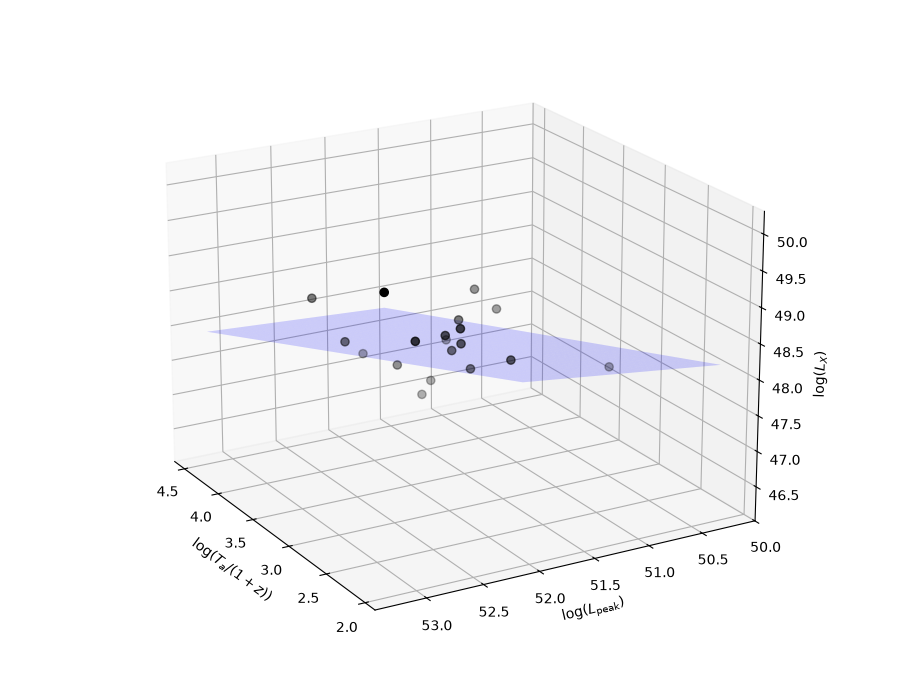}
        \caption{BNN-OHD}
    \end{subfigure}
    \hfill
    \begin{subfigure}{0.48\linewidth}
        \centering
        \includegraphics[width=\linewidth]{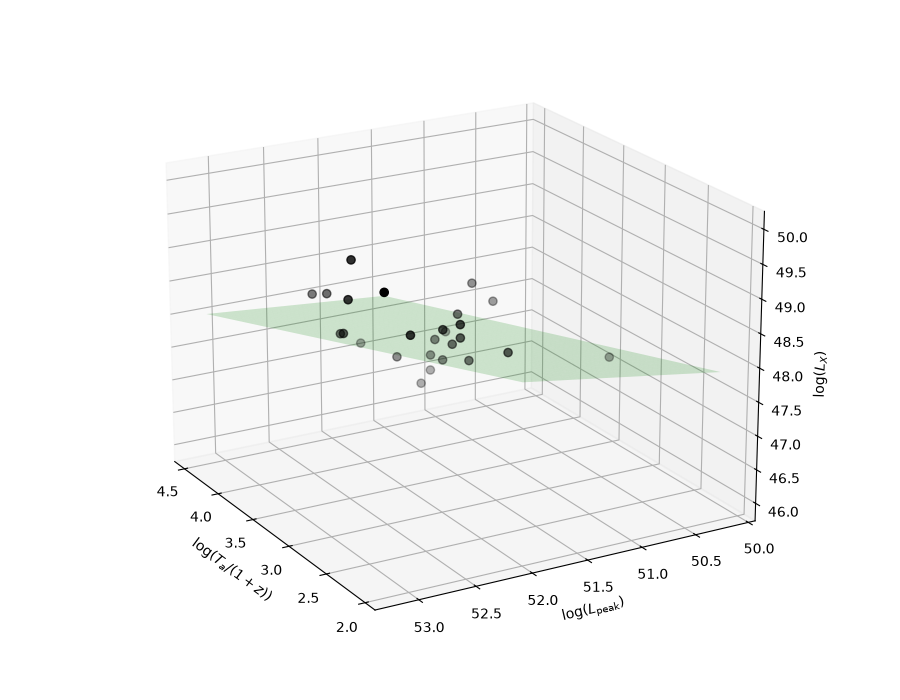}
        \caption{BNN-Pantheon}
    \end{subfigure}

    \caption{ 3D Dainotti fundamental plane for the Platinum
sub-samples
which are
calibrated
using the
(a) BNN-OHD
and
(b) BNN-Pantheon. The black markers represent the calibrated GRBs, while the colored surface indicates the best-fit 3D relation.} 
       \label{fig:combined}
\end{figure}
\noindent The three-dimensional fundamental plane correlation, also known as the 3D Dainotti correlation \cite{Dainotti2016} is given by
\begin{equation}
\log L_X
=
C_0
+
a \log T_a^{*}
+
b \log L_{\rm peak},
\end{equation}

\noindent where the peak prompt luminosity, $L_{peak}$ is-

\begin{equation}
L_{\rm peak}=4\pi d_{L}^{2}F_{\rm peak}K_{\rm prompt}
\end{equation}

\noindent while $F_{\rm peak}$ is the measured gamma-ray energy flux
at the peak of the prompt emission over a 1s interval. The likelihood can thus be obtained as :
\begin{equation}
\ln \mathcal{L}_{\rm 3D}
=
-\frac{1}{2}
\sum_{i=1}^{N}
\left[
\frac{
\left(
\log L_{X,i}
-
C_0
-
a\log T_{a,i}^{*}
-
b\log L_{{\rm peak},i}
\right)^2
}
{\sigma_i^2}
+
\ln(2\pi\sigma_i^2)
\right].
\end{equation}

\begin{figure}[H]
    \centering
    \includegraphics[width=0.8\textwidth]{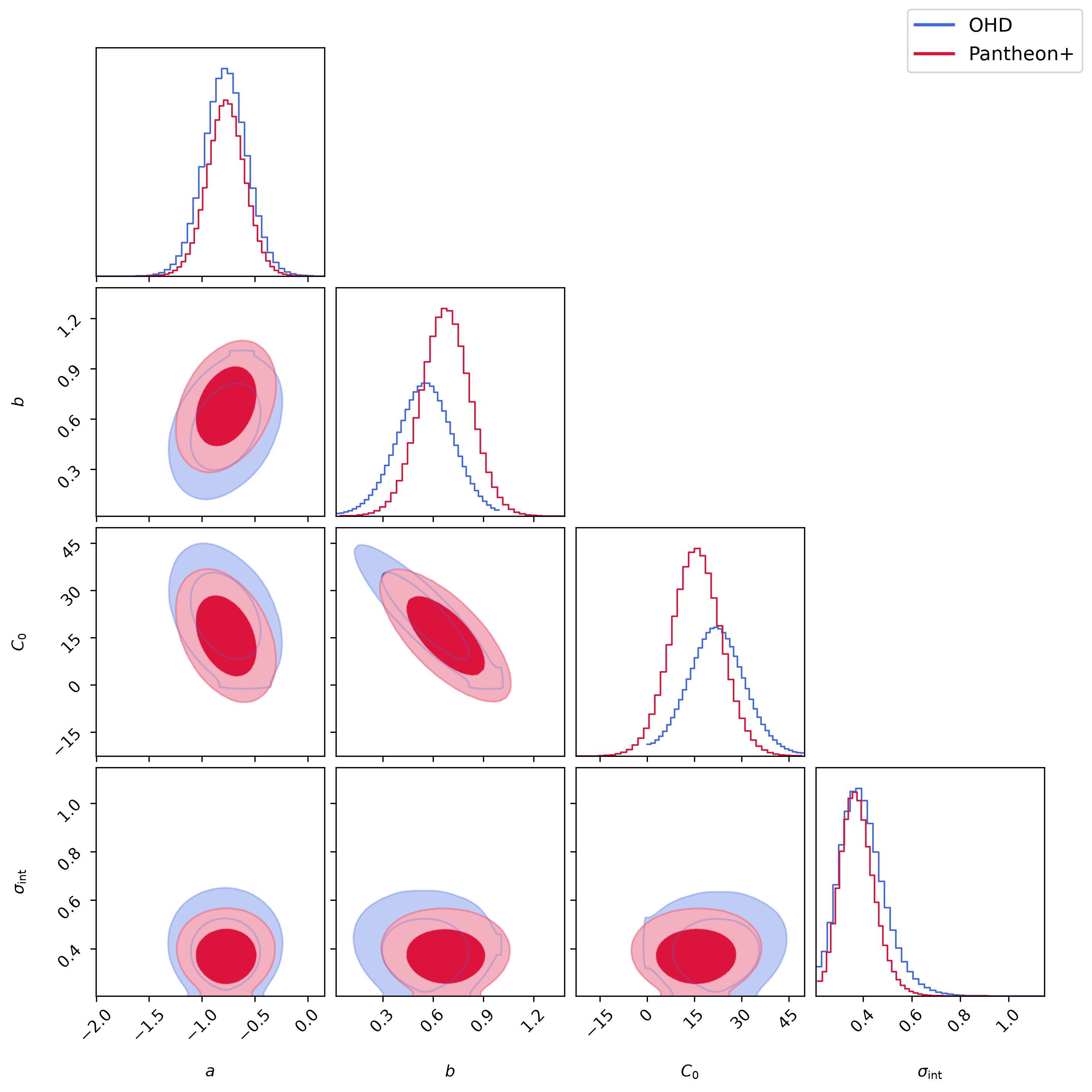}
    \caption{Corner plot of the posterior distributions for the marginalized 1$\sigma$ and 2$\sigma$ credible intervals of the parameters of the 3D Dainotti relation $(a,\,b,\,C_0,\,\sigma_{\rm int})$. Constraints obtained from calibrations using the BNN-Pantheon calibration are shown in red, while those from the BNN-OHD calibration are shown in blue.}
    \label{fig:3dplat}
\end{figure}

\noindent As per Cao et al.\cite{Cao2022g}, the GRBs favor the 3D relations strongly over the 2D relations which emphasizes the importance of the prompt phase luminosity in the plateau phase correlations. Our constraints on the parameters of the 3D Fundamental Plane are illustrated in Figure \ref{fig:3dplat}.

\section{Results}
\label{result}
\begin{sidewaystable*}[]
\centering
\label{tab:result}

\label{tab:comparison}

\begin{tabular}{l|l|l|lcccc}
\hline
Reference & Sample & Calibration & Relation & $a$ & $b$ & $C_0$ & $\sigma_{\rm int}$\\
\hline

\multirow{8}{*}{This work}

& \multirow{4}{*}{Platinum}
& \multirow{2}{*}{Pantheon+}
& 2D
& $-1.150^{+0.208}_{-0.209}$
& --
& $51.979^{+0.687}_{-0.674}$
& $0.559^{+0.097}_{-0.076}$\\
\cline{4-8}

&&
& 3D
& $-0.771^{+0.156}_{-0.157}$
& $0.675^{+0.130}_{-0.130}$
& $15.612^{+7.000}_{-6.989}$
& $0.372^{+0.066}_{-0.052}$\\
\cline{3-8}

&& \multirow{2}{*}{OHD}
& 2D
& $-1.034^{+0.203}_{-0.203}$
& --
& $51.558^{+0.669}_{-0.659}$
& $0.519^{+0.113}_{-0.083}$\\
\cline{4-8}

&&
& 3D
& $-0.775^{+0.168}_{-0.169}$
& $0.557^{+0.158}_{-0.162}$
& $21.765^{+8.721}_{-8.418}$
& $0.389^{+0.088}_{-0.065}$\\
\cline{2-8}

& \multirow{4}{*}{Narendra}
& \multirow{2}{*}{Pantheon+}
& 2D
& $-1.259^{+0.092}_{-0.091}$
& --
& $51.403^{+0.324}_{-0.326}$
& $0.889^{+0.057}_{-0.052}$\\
\cline{4-8}

&&
& 3D
& $-0.835^{+0.069}_{-0.068}$
& $0.769^{+0.056}_{-0.057}$
& $5.203^{+3.402}_{-3.397}$
& $0.585^{+0.038}_{-0.034}$\\
\cline{3-8}

&& \multirow{2}{*}{OHD}
& 2D
& $-1.255^{+0.102}_{-0.102}$
& --
& $51.379^{+0.367}_{-0.366}$
& $0.941^{+0.064}_{-0.058}$\\
\cline{4-8}

&&
& 3D
& $-0.855^{+0.069}_{-0.071}$
& $0.765^{+0.051}_{-0.056}$
& $5.498^{+3.395}_{-3.053}$
& $0.607^{+0.042}_{-0.038}$\\
\hline

\multirow{2}{*}{Mukherjee et al.}
& \multirow{2}{*}{Platinum}
& \multirow{2}{*}{Pantheon+}
& 2D
& $-1.035\pm0.162$
& --
& $51.157\pm0.535$
& $0.407\pm0.081$\\
\cline{4-8}

&&
& 3D
& $-0.830\pm0.158$
& $0.481\pm0.170$
& $25.807\pm8.978$
& $0.337\pm0.070$\\
\hline

\multirow{2}{*}{Favale et al.}
& \multirow{2}{*}{Platinum}
& \multirow{2}{*}{OHD}
& 2D
& $-1.04\pm0.16$
& --
& $51.16\pm0.53$
& $0.21^{+0.03}_{-0.05}$\\
\cline{4-8}

&&
& 3D
& $-1.00\pm0.16$
& $<0.21$
& $47.05^{+4.21}_{-1.35}$
& $0.21^{+0.03}_{-0.05}$\\
\hline

\multirow{2}{*}{Cao et al.}
& \multirow{2}{*}{Platinum}
& \multirow{2}{*}{Flat $\Lambda$CDM}
& 2D
& --
& --
& --
& --\\
\cline{4-8}

&&
& 3D
& $-0.714\pm0.104$
& $0.756\pm0.083$
& $11.52\pm4.45$
& $0.346^{+0.032}_{-0.044}$\\
\hline

\end{tabular}
\caption{Comparison of the best-fit parameters of the 2D and 3D Dainotti relations obtained for the Platinum GRB samples and Narendra Samples in this work with previous model-independent calibration studies. The table lists the fitted coefficients ($a$, $b$, and $C_{0}$) together with the intrinsic scatter ($\sigma_{\rm int}$). Results from the present BNN-based Pantheon+ and OHD calibrations are compared with the ANN-based analysis of Mukherjee et al., the Gaussian Process calibration of Favale et al., and the cosmological-model-independent calibration of Cao et al.}
\end{sidewaystable*}
The best-fit parameters obtained for the 2D Dainotti relation and 3D Fundamental Plane are summarized in Table 7. We find that
the inferred correlations are different for the Platinum and Narendra GRB samples. This may be a reflection of their different sample-selection criteria.\\

 For the Platinum sample, the 2D Dainotti relation calibrated using the BNN--Pantheon yields $a=-1.150^{+0.208}_{-0.209}$, $C_{0}=51.979^{+0.687}_{-0.674}$, and $\sigma_{\rm int}=0.559^{+0.097}_{-0.076}$, while the corresponding 3D Fundamental Plane gives $a=-0.771^{+0.156}_{-0.157}$, $b=0.675^{+0.130}_{-0.130}$, $C_{0}=15.612^{+7.000}_{-6.989}$, and $\sigma_{\rm int}=0.372^{+0.066}_{-0.052}$. Similarly, for the Narendra sample, the intrinsic scatter decreases from $\sigma_{\rm int}=0.889^{+0.057}_{-0.052}$ for the 2D relation to $\sigma_{\rm int}=0.585^{+0.038}_{-0.034}$ for the 3D Fundamental Plane. Thus, for both GRB samples, the inclusion of the prompt-emission parameter significantly tightens the correlation, consistent with previous studies of the Dainotti Fundamental Plane \cite{Cao2022s,Mukherjee:2024akt}.\\

 For both the Platinum and Narendra samples, the parameter constraints obtained using the BNN--Pantheon calibration are tighter than those obtained using the BNN--OHD calibration. For example, for the Platinum sample the uncertainty on the slope decreases from $\pm0.203$ (BNN--OHD) to approximately $\pm0.163$ (BNN--Pantheon) for the 2D relation, while for the 3D relation the uncertainty on $a$ decreases from approximately $\pm0.169$ to $\pm0.129$, with a corresponding reduction in the intrinsic scatter from $0.389^{+0.088}_{-0.065}$ to $0.349^{+0.054}_{-0.044}$. A similar improvement is observed for the Narendra sample. This is primarily because the Pantheon$+$ reconstruction extends to higher redshift ($z\lesssim2.26$), allowing a larger number of GRBs to be calibrated, whereas the OHD reconstruction is limited to approximately $z\lesssim2.0$. Consequently, the Pantheon$+$ calibration provides a larger effective calibration sample and yields more precise parameter constraints.\\

 For the Platinum sample, our results are in good agreement with previous model-independent analyses by Favale et al.~\cite{Favale:2024lgp}, Mukherjee et al.~\cite{Mukherjee:2024akt}, and Cao et al.~\cite{Cao2022s}. Although small differences are present owing to the different reconstruction techniques and calibration strategies employed, the fitted parameters are consistent within their respective uncertainties, demonstrating the robustness of the Dainotti correlations against the choice of calibration method.
\section{Conclusion and Discussion}
\label{conclu}
For GRBs to be used as reliable standard candles for cosmological studies, it is essential to employ model-independent calibration approaches to overcome the circularity problem. The resulting constraints depend strongly on the observational data used for calibration, as it determines not only the subset of GRBs that can be calibrated but also the precision of the calibration itself.\\

\noindent Our study investigates impact of use of (a) different calibration data-samples and (b) different GRBs samples. For both the GRB samples, the posterior
constraints obtained from the BNN-Pantheon predicted calibrations are tighter than those obtained by using reconstruction from BNN-OHD. This improvement is primarily a consequence of the larger calibration sample enabled by the wider redshift coverage of Pantheon+, leading to reduced parameter uncertainties. The best-fit parameters obtained from the Platinum sample are in good agreement with previous model-independent studies based on Gaussian Processes and Artificial Neural Networks. This demonstrates that the BNN reconstruction provides consistency in calibrations while also naturally propagating predictive uncertainties.\\

\noindent Our analysis also obtains that the 3D Dainotti fundamental plane exhibits a significantly smaller intrinsic scatter than the corresponding 2D correlation for the Platinum sample. This supports earlier findings that the inclusion of the prompt-emission peak luminosity provides additional physical information. This leads to a tighter standardization relation. The reconstructed parameters are also consistent with previous model-independent analyses.\\

\noindent Despite employing different reconstruction techniques for luminosity distance, the inferred slopes remain statistically consistent with previous ANN and Gaussian Process studies. This supports the relative robustness of the Dainotti relation against the choice of reconstruction methodology  provided the luminosity distances are obtained in a model-independent manner.\\

\noindent The present analysis shows that BNNs provide a robust framework for model-independent calibration of GRB luminosity relations while also propagating predictive uncertainties. However, the precision of the calibrated Dainotti relations remains limited by the size and quality of the currently available GRB samples. Future missions such as \textit{SVOM} \cite{Wei2016}, \textit{Einstein Probe} \cite{Yuan2022}, and the proposed \textit{THESEUS} \cite{Amati2021} mission are expected to substantially increase the number of well-observed GRBs with accurately measured plateau properties over a wider redshift range. Together with upcoming cosmological observations providing more precise measurements, these datasets will enable tighter model-independent calibrations of GRB luminosity correlations. The BNN framework presented here can be readily extended to these larger datasets and may provide an increasingly powerful tool for establishing GRBs as reliable standardizable candles and for probing the high-redshift universe.

\section{Acknowledgment}
N.B Aurpa gratefully acknowledges and thanks Miranda House, University of Delhi, for providing the academic environment, facilities, and support that made this work possible.
 N. Rani would like to thank IUCAA, Pune , India for providing her Vising Associateship under which a part of the work is carried out. 
\bibliographystyle{plain}
\bibliography{refs}
\end{document}